\documentclass[aps,prl,reprint, superscriptaddress,longbibliography]{revtex4-2}
\usepackage{graphicx}
\usepackage{dcolumn}
\usepackage{bm}
\usepackage{multirow}
\usepackage{verbatim}
\usepackage{xcolor}
\usepackage{setspace}
\usepackage{amsmath}
\usepackage[colorlinks=true,linkcolor=blue,citecolor=black,urlcolor=black]{hyperref} 

\newcommand{\PCO}{PdCoO$_2$}
\newcommand{\OUMU}{$\mu \Omega\, \rm{cm}$}
\newcommand{\af}{$\alpha^2 F(\omega)$}

\newcommand{\bk}{\mathbf{k}}
\newcommand{\bq}{\mathbf{q}}

\newcommand{\bK}{\mathbf{K}}

\usepackage{breakurl}
\usepackage{multirow}
\usepackage{enumitem}
\usepackage{color}
\usepackage[pagewise]{lineno}
\usepackage{tablefootnote}
\usepackage{threeparttable}
\usepackage{tabularx}
\usepackage{textcomp}
\setcitestyle{super} 

\begin{document}	
\title{Unraveling the unusually high electrical conductivity of the delafossite metal \PCO}	
\author{Xiaoping Yao}
\affiliation{School of Materials Science and Engineering, Zhejiang University, Hangzhou 310027, China}
\affiliation{Key Laboratory of 3D Micro/Nano Fabrication and Characterization of Zhejiang Province, School of Engineering, Westlake University, Hangzhou 310030, China}
\affiliation{Institute of Advanced Technology, Westlake Institute for Advanced Study, Hangzhou 310024, China}

\author{Yechen Xun}
\affiliation{Key Laboratory of 3D Micro/Nano Fabrication and Characterization of Zhejiang Province, School of Engineering, Westlake University, Hangzhou 310030, China}
\affiliation{Institute of Advanced Technology, Westlake Institute for Advanced Study, Hangzhou 310024, China}

\author{Ziye Zhu}
\affiliation{Key Laboratory of 3D Micro/Nano Fabrication and Characterization of Zhejiang Province, School of Engineering, Westlake University, Hangzhou 310030, China}
\affiliation{Institute of Advanced Technology, Westlake Institute for Advanced Study, Hangzhou 310024, China}

\author{Shu Zhao}
\affiliation{Key Laboratory of 3D Micro/Nano Fabrication and Characterization of Zhejiang Province, School of Engineering, Westlake University, Hangzhou 310030, China}
\affiliation{Institute of Advanced Technology, Westlake Institute for Advanced Study, Hangzhou 310024, China}

\author{Wenbin Li}
\email{liwenbin@westlake.edu.cn}
\affiliation{Key Laboratory of 3D Micro/Nano Fabrication and Characterization of Zhejiang Province, School of Engineering, Westlake University, Hangzhou 310030, China}
\affiliation{Institute of Advanced Technology, Westlake Institute for Advanced Study, Hangzhou 310024, China}
	\date{\today}

\begin{abstract}
  The prototypical delafossite metal \PCO\ has been the subject of
  intense interest for hosting exotic transport properties. Using
  first-principles transport calculations and theoretical modeling, we reveal
  that the high electrical conductivity of \PCO\ at room temperature
  originates from the contributions of both high Fermi velocities,
  enabled by Pd $4d_{z^2}-5s$ hybridization, and exceptionally weak
  electron-phonon coupling, which leads to a coupling
  strength ($\lambda=0.057$) that is nearly an order of magnitude
  smaller than those of common metals. The abnormally weak
  electron-phonon coupling in \PCO\ results from a low
  electronic density of states at the Fermi level, as well as the
  large and strongly facetted Fermi surface with suppressed Umklapp
  electron-phonon matrix elements. We anticipate that our work will inform the  
  design of unconventional metals with superior transport properties.
\end{abstract}

\maketitle The prototypical delafossite metal \PCO\ has attracted
considerable interest since its experimental realization for harboring
unusual electron transport
properties~\cite{Shannon_1971_1,Shannon_1971_2,Shannon_1971_3,Mackenzie2017,Daou2017}. Despite
being an oxide material, \PCO\ exhibits an exceptionally low in-plane
resistivity of 2.6~\OUMU~\cite{Hicks_2012} at room temperature, even
comparable to those of the most conducting elemental metals such as
copper and silver. Its extremely long electron mean free path, up to
60~nm at room temperature and 20~$\mu$m at low
temperature~\cite{Hicks_2012,Takatsu_2013,Mackenzie2017}, also sets a
record. Recently, there has been a renaissance of research in
\PCO\ and a number of new exotic transport phenomena have been
uncovered, including the observation of hydrodynamic electron flow and
directional ballistic transport~\cite{Moll_2016,Kikugawa_2016,
  Nandi2018, Putzke_2020, Varnavides2022, Bachmann_2022}.

Understanding the origin of \PCO’s extraordinary transport properties,
however, has been a long-standing challenge. Based on the analysis of
angle-resolved photoemission spectroscopy data, Noh et al. ascribed
\PCO's high conductivity to its high Fermi velocities, a large
two-dimensional (2D) Fermi surface and long carrier
lifetimes~\cite{Noh_2009}. Combining de Haas--van Alphen and
resistivity measurements, Hicks et al. revealed the anomalously low
contributions from electron–phonon, electron–electron, and
electron–impurity scattering to the in-plane transport of
\PCO~\cite{Hicks_2012}. A recent study indicates that the unusual high
level of crystallinity in \PCO\ is responsible for its weak
electron-impurity scattering~\cite{Sunko_2020}. However, a complete
understanding of the electron-phonon scattering, which is typically
the dominant source of resistivity near room temperature, is still
lacking~\cite{Mackenzie2017}. By carrying out fully \textit{ab initio}
transport calculations and theoretical analyses, our work addresses
the origin of the weak electron-phonon coupling and high carrier
velocities in \PCO, enabling us to provide a comprehensive
understanding of the exceptionally high electrical conductivity of
\PCO.

\begin{figure*}[t]
	\centering
	\includegraphics[width=0.95\textwidth]{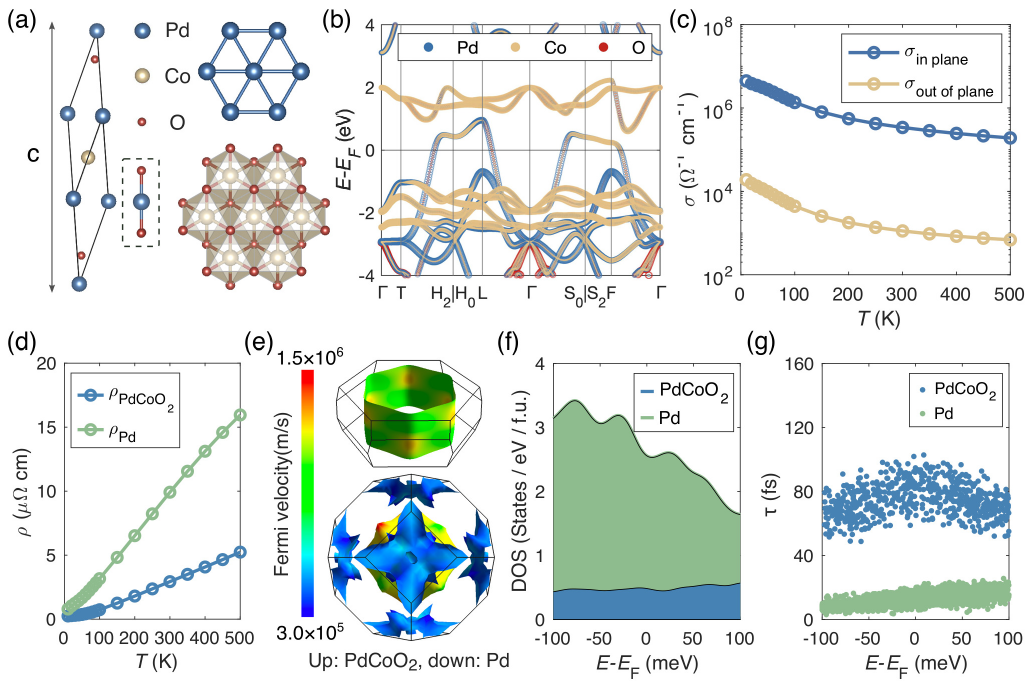}
	\caption{Atomistic models and electronic transport properties
          of \PCO. (\textbf{a}) Primitive cell of \PCO, triangular
          lattice of Pd, and edge-sharing CoO$_6$ octahedra. The
          O-Pd-O dumbbell in the dashed box demonstrates the twofold,
          linearly coordinated environment of Pd. (\textbf{b})
          Electronic band structure of \PCO\ calculated from density
          functional theory. The corresponding reciprocal space
          path is illustrated in Fig.~S1b. The atomic orbital
          characters of the bands are indicated using open circles of
          different colors. (\textbf{c}) Comparison of the intrinsic electrical
          conductivity along the in-plane and out-of-plane directions
          in \PCO, calculated by \textit{ab initio} Boltzmann
          transport equation. (\textbf{d}) Comparison between the in-plane
          resistivities of \PCO\ and Pd. (\textbf{e}) Fermi velocities
          of \PCO\ (upper panel) and Pd (bottom panel) projected on
          the corresponding Fermi surfaces. (\textbf{f}) Comparison of
          the carrier density of states (DOS) per formula unit near
          the Fermi level of \PCO\ and Pd. The zero of the horizontal
          axis represents the Fermi level. (\textbf{g}) Comparison of
          the carrier lifetimes of \PCO\ and Pd at 300K.}
\label{fig:fig1}
\end{figure*}

\begin{figure*}[t]
  \centering
    \includegraphics[width=0.98\textwidth]{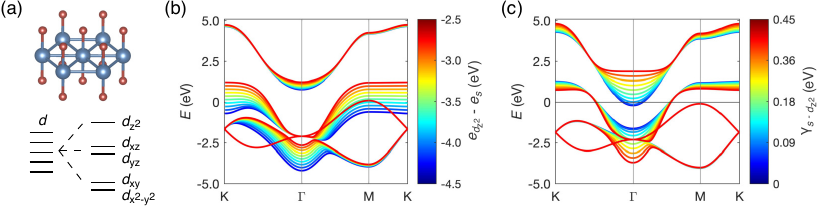}
    \caption{Origin of the high Fermi velocities in \PCO.
      (\textbf{a}) The coordinate environment around Pd in \PCO\ and
      the resulting unusual uniaxial crystal field. Each Pd atom (blue
      ball) is linearly coordinated with two O atoms (red balls),
      leading to $d$-orbital splitting, with the $d_{z^2}$ orbital
      occupying the highest level. (\textbf{b},\textbf{c}) Evolution
      of the tight-binding (TB) bandstructure of a corresponding Pd
      sheet with TB parameters: (\textbf{b}) The on-site energy
      difference between the $5s$ and $4d_{z^2}$ orbitals
      $e_{d_{z^2}}-e_{s}$, and (\textbf{c}) the hopping integral
      between the $5s$ and $4d_{z^2}$ orbitals $\gamma_{s-d_{z^2}}$.}
    \label{fig:TB}
\end{figure*}
  
\PCO\ crystallizes in rhombohedral delafossite-type structure, with
$\textit{R}\bar{3}m$ space group symmetry. It comprises alternative
layers of Pd and edge-shared CoO$_6$ octahedra flattened with respect
to the $c$-axis, each of which forms an individual triangular lattice,
as shown in Fig.~\ref{fig:fig1}a. These two alternating layers are
connected by the O-Pd-O dumbbells, where Pd is linearly coordinated to
two oxygen atoms (see Fig.~S1a) and formally in a monovalent
state~\cite{noh2009orbital,Tanaka_1998,Noh_2009}.

First, we summarize the first-principles electronic structure of
\PCO\ in Fig.~\ref{fig:fig1}b. Only a single steep band crosses the
Fermi level. The strongest contribution near the Fermi level was found
to come from Pd. Further orbital decomposition reveals that the states
near the Fermi level mainly derive from the $\{d_{z^2}, d_{x^2-y^2},
d_{xy}\}$ orbitals of Pd, and to a lesser extent, the $s$ and $p$
orbitals of Pd (see Fig. S2 and Fig. S3). The associated Fermi surface
(see Fig.~\ref{fig:fig1}e) is highly two-dimensional and nearly
hexagonal~\cite{Hicks_2012}.

We obtain the intrinsic, phonon-limited electrical conductivity and
resistivity of \PCO\ and, for comparison, metal Pd, by solving the
linearized Boltzmann transport equation (BTE) with fully \textit{ab
  initio} electron-phonon coupling~\cite{Ponce2020, Giustino2017} (see
Supplementary Method). The results are shown in
Fig.~\ref{fig:fig1}(c,d). The calculated room-temperature in-plane
resistivity of \PCO, 2.9~\OUMU, demonstrates a close agreement with
the experimental value of 2.6~\OUMU~\cite{Hicks_2012}. Besides,
\PCO\ exhibits a strong anisotropy of conductivity, with the in-plane
conductivity several hundred times higher than that in the
perpendicular direction. This prevailing in-plane character is
consistent with its quasi-2D structure. Compared with the pure metal
Pd, \PCO\ has significantly (more than a factor of three) lower
resistivities in all temperature regimes.

To unveil the origin of the exceptionally low in-plane resistivity
found in \PCO, we have investigated the key parameters that influence
the conductivity of a metal, including Fermi velocity, charge carrier
density, and carrier lifetime, and further made a comparison with
Pd. The Fermi velocities ($v_F$) of \PCO\ and Pd are projected onto
their respective Fermi surface in Fig.~\ref{fig:fig1}e, from which it
is seen that Pd can have both high and low $v_F$ (see also the band
structure of Pd in Fig.~S4), whereas \PCO\ has relatively uniform and
high $v_F$. It can be deduced that, on average, \PCO\ has higher $v_F$
compared with Pd, which contributes to its higher conductivity. Thus,
a key issue in understanding \PCO's high conductivity is the origin of
its high Fermi velocities. We will return to this in light of our
tight-binding (TB) model. In Fig.~\ref{fig:fig1}f, we compare the
carrier density of states (DOS) per unit cell near the Fermi level in
\PCO\ and Pd. It is seen that the DOS of \PCO\ is, on average, nearly
five times smaller than that of Pd within the 100~meV range around
their corresponding Fermi level. Adding the fact that the unit-cell
volume of \PCO\ is nearly 2.7 times larger than that of Pd, it can be
deduced that the charge carrier density in \PCO\ is significantly
lower than that in Pd.  In Fig.~\ref{fig:fig1}g, we further compare
the carrier lifetimes of \PCO\ and Pd at 300~K, for energies within
the 100 meV range around their corresponding Fermi level. The result
demonstrates that \PCO's carrier lifetimes (about 80~fs) are almost
one order of magnitude larger than that of Pd. As a result of the high
Fermi velocities and long carrier lifetimes, the in-plane mean free
path of \PCO\ has large values around 60 nm at room temperature
(Fig.~S5). On the basis of the above results, it can be concluded that
the reason why \PCO\ has remarkably high conductivity is that it
possesses both highly mobile and extremely long-lived electrons.

We now return to the subject of high Fermi velocities in \PCO, whose
origin has been in debate, especially on the role played by the Pd
$5s$ states~\cite{Hicks_2012, Shannon_1971_3, Tanaka_1998,
  Seshadri1998, Eyert2008, Kim2009fermi, Ong2010, Usui_2019}. Here we
illustrate how the high Fermi velocities arise in a TB
model. Specifically, we developed a minimal four-band ($s$ $d_{z^2}$
$d_{x^2-y^2}$ $d_{xy}$) TB model involving only the nearest-neighbor
Pd-Pd hoppings on a simple bare Pd triangular-lattice sheet, with Pd
in a monovalent state. The detail of the TB model and can be
found in Supplementary Note 1, and the DFT-calculated and TB-fitted
band structures are shown in Fig.~S6 and Fig.~S7, respectively. We
then examine the bandstructure evolution by varying the TB parameters,
as shown in Fig.~\ref{fig:TB}. With the on-site energy level of the
$4d_{z^2}$ orbital increased, the band composed mainly by $4d_{z^2}$
shifts to higher in energy and becomes the only conduction band. The
slope of this conduction band (and hence the Fermi velocities) can be
effectively increased by increasing the hopping integral between the
$5s$ and $4d_{z^2}$ orbitals.

On the basis of the TB model, the high Fermi velocities of \PCO\ can
be ascribed to the unusual crystal field splitting provided by the
unique twofold, linearly coordinated Pd in \PCO, which splits the $4d$
orbitals of Pd into three degenerate subsets, $\{d_{z^2}\}$,
$\{d_{xz}, d_{yz}\}$, and $\{d_{xy}, d_{x^2-y^2}\}$, with the
$\{d_{z^2}\}$ subset occupying the highest energy, as illustrated in
Fig.~\ref{fig:TB}a. The crystal field environment and the reduced
energy gap between the $5s$ and $4d_{z^2}$ orbitals also lead to
enhanced $5s - 4d_{z^2}$ hybridization~\cite{Shannon_1971_3}. Together
with the monovalent state of Pd, these conditions result in a single,
highly dispersive band that crosses the Fermi level.  This model is
further corroborated by an artificially constructed crystal model of
PdF that has a similar structure as \PCO\ (see Fig.~S8a), in which F
atoms replace the CoO$_2$ layers and serve as electron acceptor, thus
allowing Pd to remain in a monovalent state. By controlling the
crystal field of Pd through the stretching or compression of the Pd-F
distance, we are able to significantly tune the Fermi velocities of
the system (see Fig.~S8).

The other factor contributing to \PCO's high conductivity, as
mentioned above, is its exceptionally long carrier lifetimes $\tau$,
which motivates us to query the reason behind. For metals, when the
temperature is above the Debye temperature, the carrier lifetime can
be estimated by $\displaystyle \tau = \hbar \left( 2\pi
  k_BT \lambda\right)^{-1}$, where $\lambda$ is the electron-phonon coupling
strength~\cite{Allen78, Grimvall1981}. $\lambda$ is computed from the
Eliashberg spectral function \af\ as $\displaystyle \lambda = 2 \int_0^{\infty}
d\omega \, \omega^{-1} \alpha^2 F(\omega)$,
where $\alpha^2 F(\omega)$ is given by
\begin{align}
  \alpha^2 F(\omega) = & \frac{1}{N_{\textrm{F}}} \sum_{mn\nu}
  \int \frac{d\bk d\bk'}{\Omega_{\textrm{BZ}}^2} |g_{mn\nu}(\bk,
  \bk')|^2 \delta(\varepsilon_{n\bk}-\varepsilon_{\textrm{F}})
  \nonumber \\
  & \times \delta(\varepsilon_{m\bk'}-\varepsilon_{\textrm{F}})
  \delta(\hbar \omega - \hbar \omega_{\bk'-\bk, \nu}).
\end{align}
Here, $N_{\textrm{F}}$ represents the DOS per spin per unit cell at
the Fermi level~$\varepsilon_{\textrm{F}}$. The integrations are over
the Brillouin zone of volume $\Omega_{\textrm{BZ}}$.  $g_{mn\nu}(\bk,
\bk') = \langle \psi_{m\bk'} \vert \Delta_{\bq\nu}V_{\textrm{scf}}
\vert \psi_{n\bk}\rangle $ is the electron-phonon matrix element for a
Bloch wave with a band index $n$ and momentum $\bk$ scattered to
another state $m\bk'$, due to the variation of the self-consistent
potential $V_{\textrm{scf}}$ induced by a phonon with momentum $\bq$
and branch index $\nu$. Conservation of crystal momentum requires that
$\bk'-\bk = \bq + \mathbf{G}$, with $\mathbf{G} = 0$ and
$\mathbf{G}\neq 0$ corresponding to normal and Umklapp scattering,
respectively.

\begin{figure*}[t]
    \centering
      \includegraphics[width=1\textwidth]{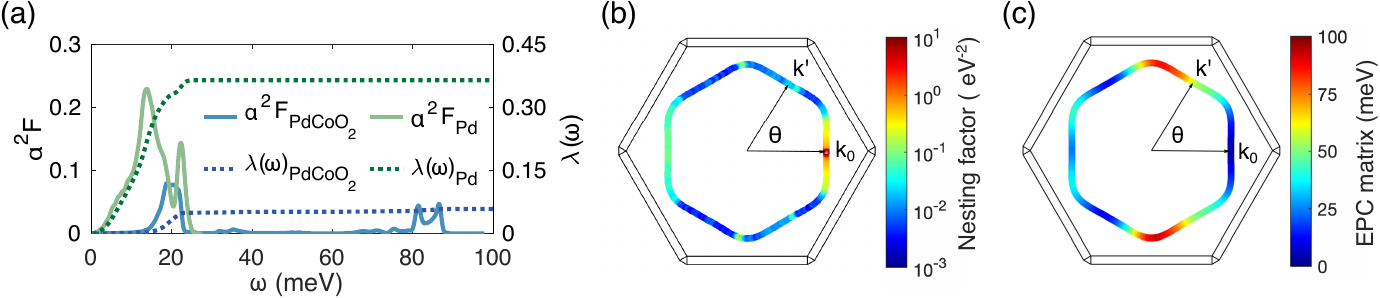}
      \caption{Electron-phonon coupling in \PCO\ and Pd.  (\textbf{a})
        Comparison of the Eliashberg spectral function \af\ and the
        associated electron-phonon coupling strength $\lambda$ between
        \PCO\ and Pd, with the blue and green lines representing
        \PCO\ and Pd, respectively. (\textbf{b}) The nesting factor
        ($\xi_{\bk'-\bk_0}$) along a cross section ($k_z = 0$) of the Fermi surface
        in \PCO, with the initial electronic state fixed at $\bk_0$
        and the final located at $\bk'$ along the Fermi surface.  The
        angle $\theta$ represents the direction of $\bk'$ with respect
        to $\bk_0$. (\textbf{c}) Sum of the absolute values
        of the acoustic electron-phonon matrix elements between $\bk_0$ and
        $\bk'$ along the Fermi surface section in \PCO.}
      \label{fig:elph}
\end{figure*}

In Fig.~\ref{fig:elph}a, we compare the calculated $\lambda$ and
\af\ of \PCO\ and Pd. Surprisingly, the coupling strength of
\PCO\ ($\lambda = 0.057$) is nearly an order of magnitude smaller than
that of Pd ($\lambda = 0.36$). It is exceptionally small even compared
to other highly conducting metals such as Cu ($\lambda =
0.14$)~\cite{Savrasov1996, Grimvall1981}. This is also reflected in
the \af\ of \PCO, whose value is rather small in the low-frequency
range, before a peak emerges at a relatively high acoustic-phonon
frequency of $\sim$20~meV (see also Fig.~S9). The peaks in the \af\ of
\PCO\ at frequencies above 80~meV come from high-frequency optical phonons that
correspond to the Pd-O bond stretching/compression
modes~\cite{Takatsu2007, Homes2019}, but their contribution to
$\lambda$ is relatively minor ($\sim$12\%, see also Fig.~S10). Using
the computed $\lambda$ value, the carrier lifetime of \PCO\ at 300~K
is estimated to be $\sim$71~fs, in close agreement with the \textit{ab
  initio} values that average around 80~fs. We have also computed the
transport Eliashberg spectral function $\alpha^2_{\textrm{tr}}
F(\omega)$ and transport coupling strength $\lambda_{\textrm{tr}}$ of
\PCO\ and Pd in Fig.~S11, with very similar results
($\lambda_{\textrm{tr}} = 0.056$ for \PCO).

To understand the abnormally small $\lambda$ of \PCO\, we note that
$\lambda$ can be written as~\cite{Grimvall1981}:
\begin{equation}
 \lambda = 2N_{\textrm{F}} \Bigl\langle\Bigl\langle \sum_{\nu}
 \frac{|g_{mn\nu}(\bk, \bk')|^2}{\hbar \omega_{\bk'-\bk, \nu}}
 \Bigl\rangle\Bigl\rangle_{\textrm{FS}},
 \label{eq:eq2}
\end{equation} 
where $\langle\langle \cdots \rangle \rangle_{\textrm{FS}}$ denotes a
double average over the electronic states $m\bk$ and $n\bk'$ on the
Fermi surface. Hence, a
key factor that results in the exceptionally small $\lambda$ of
\PCO\ is its small DOS $N_{\textrm{F}}$, which originates from its single, highly
dispersive band that crosses the Fermi level.

However, DOS alone cannot fully explain \PCO's $\lambda$ value. In
particular, if we focus on the acoustic-phonon contribution to
$\lambda$ by setting the upper limit of the frequency integration of
$\omega^{-1} \alpha^2 F(\omega)$ to $\omega_{\textrm{max}}= 30$~meV
for \PCO, the calculated $\lambda$ of \PCO\ is only
$\lambda_{\textrm{PdCoO}_2}^{\textrm{acoustic}} = 0.05$. This leads to
$\lambda_{\textrm{Pd}}/\lambda_{\textrm{PdCoO}_2}^{\textrm{acoustic}}
\approx 7.2$, which is much larger than the corresponding DOS ratio
near the Fermi level
($N_{\textrm{F}}(\textrm{Pd})/N_{\textrm{F}}(\textrm{PdCoO}_2)\approx
5$). The Fermi surface average of $|g|^2/\omega$, therefore, must also
play an important role. To elucidate its contribution, we calculate
the zeroth-moment frequency integration of \af, denoted by
$\displaystyle \kappa(\omega) = 2\int_0^{\omega} d\omega' \, \alpha^2
F(\omega')$.  $\kappa$ provides information regarding the Fermi
surface average of $|g|^2$: $\kappa = 2N_{\textrm{F}}
\Bigl\langle\Bigl\langle \sum_{\nu} |g_{\nu}(\bk, \bk')|^2/\hbar
\Bigl\rangle\Bigl\rangle_{\textrm{FS}}$. For phonon frequency up to
$\omega_{\textrm{max}}$, we find that
$\kappa_{\textrm{Pd}}/\kappa_{\textrm{PdCoO}_2}^{\textrm{acoustic}}
\approx 4.9$, which is close to the DOS ratio. Thus, the Fermi surface
average of the $|g|^2$, summing over the acoustic phonon branches, are
similar between \PCO\ and Pd. What contributes to the much smaller
$\lambda$ value of \PCO, in addition to its small DOS, is that the
electron-phonon scattering in \PCO\ disproportionately occurs in the
higher-frequency acoustic phonon region, as can be seen from the
\af\ data in Fig.~\ref{fig:elph}a. This behavior can be understood
given that \PCO\ has a large, open, and highly facetted Fermi surface,
which gives rise to Fermi surface nesting between the parallel sheets
of the nearly hexagonal surface. This is quantified in
Fig.~\ref{fig:elph}b, where we plot the nesting factor $\xi_{\bk'-\bk}
= \sum_{nm} \int_{\textrm{BZ}} \frac{d\bk}{\Omega_{\textrm{BZ}}}
\delta(\varepsilon_{n\bk}-\varepsilon_{\textrm{F}})
\delta(\varepsilon_{m\bk'}-\varepsilon_{\textrm{F}})$ for different
$\bk'-\bk$ values. The frequency of the phonons that corresponds to
the nesting vector is around 20~meV (Fig.~S12), in consistent with the
main peak of \af\ of \PCO\ in Fig.~\ref{fig:elph}a.

Another perspective to understand the contribution of $|g|^2/\omega$
to $\lambda$ is that the matrix elements $|g|$ corresponding to the
Umklapp backscattering are suppressed in \PCO. This is indeed
supported by our \textit{ab initio} calculations of the
electron-phonon matrix elements $\sum_{\nu}|g_{\nu}(\bk, \bk')|$ for
the acoustic phonon branches in Fig.~\ref{fig:elph}c. When $\bK =
\bk'-\bk$ enters into the Umklapp scattering regime, which roughly
corresponds to $\theta = \pi/2$ in Fig.~\ref{fig:elph}c,
$\sum_{\nu}|g_{\nu}(\bk,\bk')|$ becomes lower in magnitude (see also
Fig.~S13). Meanwhile, the phonon frequency $\omega$ does not exhibit a
significant variation (see Fig.~S12).  We verify that the change in
$|g|$ is not directly correlated with the variation in the overlap of
the lattice periodic parts of the wavefunctions $\langle u_{\bk'} |
u_{\bk} \rangle$ (due to Fermi surface orbital-momentum
locking~\cite{Usui_2019}, see Fig.~S14).  Instead, the result is more
consistent with the screening model of electron-phonon
interaction~\cite{Giustino2017}, in which the electronic screening of
the phonon-induced change in nucleus potential plays a pivotal role in
determining the strength of electron-phonon coupling. In the simplest
version of the model, the electron-phonon matrix elements can be written
as~\cite{Giustino2017}:
\begin{equation}
g_{\nu}(\bk, \bk') = -i \sqrt{\frac{\hbar}{2M\omega_{\bK, \nu}}} \bK \cdot
\mathbf{e}_{\nu}(\bK) \frac{V(\bK)}{\epsilon(\bK)},
\label{eq:eq3}
\end{equation}
where $\mathbf{e}_{\nu}(\bK)$ denotes the polarization vector of the
phonon mode $\bK\nu$, $V(\bK) = -4\pi e^2/|\bK|^2$ the Fourier
transform of the electron-nucleus potential energy, $\epsilon(\bK)$
the dielectric function, and $M$ the nucleus mass. If neglecting the
exchange-correlation contribution, $\epsilon(\bK)$ can be approximated
by the Lindhard dielectric function $\epsilon(\bK) = 1 +
q_{\textrm{LH}}^2(\bK)/|\bK|^2$, where $q_{\textrm{LH}}^2(\bK) = -4\pi
e^2 \chi_0(\bK)$, with $\chi_0(\bK)$ being the static bare
susceptibility: $\chi_0(\bK) = \frac{2}{N_{\bk} \Omega} \sum_{\bk}
\frac{f_{\bk + \bK} - f_{\bk}}{\varepsilon_{\bk + \bK} -
  \varepsilon_{\bk}}$. The screened potential energy
$V(\bK)/\epsilon(\bK)$ in Eq.~\ref{eq:eq3} can then be written as
$-4\pi e^2 \left[\bK^2 + q_{\textrm{LH}}^2(\bK)\right]^{-1}$. In
Fig.~S15 and S16, we have further ruled out the polarization factor
$\bK \cdot \mathbf{e}_{\nu}(\bK)$ in Eq.~\ref{eq:eq3} as the source of
the matrix-element suppression. Instead, Fig.~S17 indicates that, due
to the Fermi surface nesting, $-\chi_0(\bK)$ and
$q_{\textrm{LH}}(\bK)$ are significantly larger in the backscattering
directions, which, in combination with the large $|\bK|$ involved,
leads to much smaller $\left[\bK^2 +
  q_{\textrm{LH}}^2(\bK)\right]^{-1}$ and hence
weakened $V(\bK)/\epsilon(\bK)$.  We thus conclude that the electron-phonon
backscattering in \PCO, while favoured by the Fermi surface nesting,
is simultaneously associated with weak electron-phonon matrix
elements, which are most likely caused by the enhanced electronic
screening and a large Umklapp momentum transfer.

In summary, our comprehensive study unravels that the extremely high
in-plane conductivity of \PCO\ originates from the combination of both
high Fermi velocity and exceptional weak electron-phonon coupling
($\lambda = 0.057$). The later is a result of the relatively low
electronic density of states at the Fermi level, as well as the 
large, strongly facetted Fermi surface that leads to Fermi surface
nesting and suppressed Umklapp electron-phonon matrix elements. Our
work provides significant insights into the electron-phonon coupling
in delafossite metals and informs the design of unconventional metals
with ultralong carrier lifetimes and superior room-temperature
conductivities.

\begin{acknowledgements}
We gratefully acknowledge the support by NSFC under Project
No. 62004172 and Research Center for Industries of the Future at
Westlake University under Award No. WU2022C041. W.L. thanks
Prof. Qiyang Lu for drawing his attention to the delafossite metal
system and for helpful discussions. W.L. is also indebted to
Prof. Andrew Mackenzie for discussions and very helpful comments on
the manuscript. The authors thank Jiaming Hu for useful discussions
and the Westlake University HPC Center for computational support.
\end{acknowledgements}

\end{document}